\documentclass[aps,prb,twocolumn,groupedaddress]{revtex4}
\usepackage{graphics,epsfig}
\usepackage{amssymb}
\newcommand{\beq}{\begin{equation}}
\newcommand{\eeq}{\end{equation}}
\newcommand{\pdag}{{\phantom{\dagger}}}
\newcommand{\Jpar}{J_{\|}}
\newcommand{\Jperp}{J_{\bot}}
\begin{document}


\title{Violation of the Fluctuation-Dissipation Theorem 
and Heating Effects in the Time-Dependent Kondo Model}


\author{Dmitry Lobaskin}
\author{Stefan Kehrein}
\affiliation{Theoretische Physik III -- Elektronische Korrelationen und
Magnetismus, Universit{\"a}t Augsburg, 86135 Augsburg, Germany}


\date{\today}

\begin{abstract}
 The fluctuation-dissipation theorem (FDT) plays a fundamental role in 
understanding quantum many-body problems. However, its applicability 
is limited to equilibrium systems and it
 does in general not hold in nonequilibrium situations. This violation of the FDT
 is an important tool for studying nonequilibrium physics. 
 In this paper we present results for the violation of the FDT in the Kondo model where the impurity spin is frozen for all negative times, and set free to relax at positive times. We derive exact analytical 
 results at the Toulouse point, and results within a controlled approximation
 in the Kondo limit, which allow us to study the FDT violation on all time scales.
 A measure of the FDT violation is provided by the {\em effective temperature},
 which shows initial heating effects after switching on the perturbation, and
 then exponential cooling to zero temperature as the Kondo system reaches
 equilibrium.
\end{abstract}


\maketitle

\section{Introduction}
The fluctuation-dissipation theorem (FDT)\cite{CW} is of fundamental importance for
the theoretical understanding of many-body problems. It establishes a relation
between the equilibrium properties of a system and its response to
an external perturbation. In nonequilibrium situations this powerful tool is 
in general not available: typical nonequilibrium situations are e.g.\
systems prepared in an excited state, or systems driven 
into an excited state by pumping energy into them. Since such
nonequilibrium systems occur everywhere in nature, the investigation
of nonequilibrium many-body physics has become one of the key challenges
of modern theoretical physics. The violation of the FDT in a 
nonequilibrium system plays an important role in such studies as it
characterizes ``how far" the system is driven out of equilibrium. 

Most widely investigated in this context are glassy systems, that is 
systems with a very long relaxation time compared to the typical time scale of measurements. Due to the long relaxation times it is experimentally possible to measure the deviation from the FDT, i.e. to
study the {\em ageing effects}: one observes a relaxation of the
nonequilibrium initial state towards equilibrium. For a review of this field see 
Refs.~\onlinecite{HertzFisher,Calabrese_review}.

However, these are classical systems at finite temperature and therefore the classical limit of the FDT is studied. Nonequilibrium zero temperature quantum systems provide
a very different limit which has been studied very little in the literature
(however, see Ref.~\onlinecite{Mauger}). 
This is one of the main motivations for our work which looks at the
FDT violation in the time-dependent Kondo model at zero temperature. 
Time-dependence is here introduced by freezing the impurity spin at 
negative times, and then allowing it to relax at positive times. 
Besides being of fundamental theoretical importance as the paradigm
for strong-coupling impurity physics in condensed matter theory, Kondo
physics is also experimentally realizable in quantum dots. The Kondo effect
has been observed in quantum dot experiments\cite{Kondo_dots}, and
time-dependent switching of the gate potential amounts to a 
realization of the time-dependent Kondo model\cite{Nordlander} which should
be possible in future experiments.  

Our calculations here are based on recent work on the time-dependent
Kondo model with exact analytical results for the Toulouse point and 
results in a controlled approximation in the experimentally relevant
Kondo limit.\cite{article} We will see that the FDT is maximally
violated at intermediate time scales of order the inverse Kondo
temperature: the effective temperature becomes of order the
Kondo temperature due to heating of the conduction band electrons
by the formation of the Kondo singlet. The system then relaxes 
towards equilibrium and the FDT becomes fulfilled exponentially fast 
at larger times.

\section{Fluctuation-Dissipation Theorem}
Consider an observable $A$ which is coupled linearly
to a time-dependent external field $h(t)$. The Hamiltonian of the system 
is  then given by
\beq
H=H_0-h(t)\,A \ ,
\eeq 
where $H_0$ is the unperturbed part of the Hamiltonian. 
The generalized susceptibility (or {\em response})  $R(t,t')$ of the observable 
$A$ at time $t$ to
the external small perturbation $h(t')$ at time~$t'$ is defined as 
\beq
R(t,t') =\left.\frac{\delta \langle \overline{A(t)} \rangle}{\delta h(t')}\right|_{h=0} \ .
\label{response}
\eeq
Here $\langle \overline{A(t)} \rangle \equiv \langle A(t) \rangle_h - \langle A(t) \rangle_0$ is the deviation of the expectation value from its equilibrium value.
If the system is in equilibrium before its perturbation by the field $h$, 
then $R(t,t')$ depends only on the time difference $\tau = t-t'$. One introduces 
the Fourier transform of $R(\tau)$
\beq
R(\omega) = \int_0^{\infty} R(\tau) e^{i\omega \tau} d\tau \ ,
\eeq
where the integration runs only over positive times as a consequence of causality.
A simple calculation shows that the imaginary part of $R(\omega)$ is proportional to the energy dissipated by the system for a small periodic perturbation with frequency~$\omega$ 
(see, for example, Ref.~\onlinecite{LL}). Thus the response function determines the dissipation properties of the equilibrium system.

For calculating the response function one defines the
two-time correlation function
\beq
C_{A,A}(t,t')\equiv \left\langle A(t)A(t')\right\rangle =\frac 1ZTr\left[
A(t)A(t')\rho \right] \ ,
\eeq
with the operators in the Heisenberg picture
\beq
A(t)\equiv \exp \left( iHt \right) A(0)\exp \left( -iHt \right) \ .
\eeq
The trace runs over all the states in the Hilbert space, 
$\rho$ is the density matrix and $Z$ the partition function. 
Symmetrized and antisymmetrized correlation functions $C_{\left\{ A,A\right\} }(t,t')$, $C_{\left[ A,A\right] }(t,t')$ are defined in the same way
\begin{eqnarray}
C_{\left\{ A,A\right\} }(t,t')&\equiv & \frac 12\left\langle \left\{
A(t),A(t')\right\} \right\rangle 
\label{cor_sym} \\
C_{\left[ A,A\right] }(t,t')&\equiv & \frac 12\left\langle \left[
A(t),A(t')\right] \right\rangle 
\label{cor_antisym} \ .
\end{eqnarray}
The cumulant of the symmetrized correlation function is 
\beq
C^{\rm (cum)}_{\{A,A\}}(t,t')\equiv C_{\{A,A\}}(t,t')
-\langle A(t)\rangle\,\langle A(t')\rangle \ .
\eeq
In the framework of linear response theory (that is for small perturbations)
one then proves the famous Kubo formula\cite{Kubo}
\beq
R(t,t')=2i \theta (t-t')C_{\left[ A,A\right] }(t,t') \ .
\eeq
Here the time dependence of all operators is given by the unperturbed part of the Hamiltonian $H_0$.
Since in equilibrium all correlation functions depend only on the time
difference $\tau=t-t'$, one defines their Fourier transform with respect to~$\tau$ 
\beq
C^{\rm (cum)}_{\{A,A\}}(\omega)=\int_{-\infty}^{\infty} 
C^{\rm (cum)}_{\{A,A\}}(\tau)\, e^{i\omega \tau}  d\tau
\ .
\label{fourier_full}
\eeq
If the initial state is the equilibrium state for a given temperature, $R(\omega)$ and $C^{\rm (cum)}_{\{A,A\}}(\omega)$ are related by the famous Callen-Welton relation\cite{CW}, which is also known as $the$ Fluctuation-Dissipation theorem
\beq
{\rm Im}\, R(\omega) = \tanh\left(\frac{\beta\omega}{2}\right) C^{\rm (cum)}_{\{A,A\}}(\omega) \ .
\label{FDT}
\eeq
Here $\beta$ is the inverse temperature. For $T=0$ Eq.~(\ref{FDT}) reads 
\beq
{\rm Im}\, R(\omega) = {\rm sgn}\left(\omega\right) 
C^{\rm (cum)}_{\{A,A\}}(\omega) \ .
\label{FDT_zero}
\eeq
Eqs.~(\ref{FDT}) and (\ref{FDT_zero})\cite{footnote_cumulant} 
relate dissipation with equilibrium fluctuations, which is the
fundamental content of the FDT.

\section{FDT Violation in Nonequilibrium} 
Let us recapitulate why the FDT (\ref{FDT}) in general will not hold
in quantum nonequilibrium systems. 
We will only consider the zero temperature case since it brings out
the quantum effects most clearly; the generalization to nonzero temperatures 
is straightforward. 

We first consider how a typical experiment is actually performed: the
system in prepared in some initial state at time~$t=0$ (not necessarily
its ground state) and then evolves according to its Hamiltonian. 
A response measurement
is then done by applying the external field after a
{\em waiting time} $t_w>0$, and the response to this is measured a
time difference~$\tau$ later. The Fourier transform with respect to
(positive) time difference will then in general depend on the 
waiting time~$t_w$
\beq
R(\omega, t_w) = \int_0^{\infty} R(t_w+\tau,t_w)\, e^{i\omega \tau} d\tau \ .
\label{Romegatw}
\eeq
Likewise in an experimental measurement of the correlation function
the first measurement of the observable will be performed after
the waiting time~$t_w$, and then at time $t_w+\tau$ the second
measurement follows. From the experimental point of view this
leads again to a one-sided Fourier transform
\beq
C^{\rm (cum)}_{\{A,A\}}(\omega,t_w)=2 \int_{0}^{\infty} 
C^{\rm (cum)}_{\{A,A\}}(t_w+\tau,t_w)\, \cos (\omega \tau) d\tau \ .
\label{Comegatw}
\eeq
If the system is prepared in its ground state, or if the system
equilibrates into its ground state for sufficiently long waiting
time $t_w\rightarrow\infty$, then we can replace this one-sided
Fourier transform by the symmetric version and arrive at the
conventional equilibrium definition (\ref{fourier_full}).

However, for a nonequilibrium preparation at $t=0$ 
Eqs.~(\ref{Romegatw}) and (\ref{Comegatw}) are the suitable
starting point for the discussion of the FDT (\ref{FDT_zero}).
Let us therefore look at the FDT in the
framework of (\ref{Romegatw}) and (\ref{Comegatw}).
We follow the standard derivation\cite{LL}  and introduce
a complete set of eigenstates $|n\rangle$ of the 
Hamiltonian $H$, $H|n\rangle=E_n|n\rangle$. The matrix elements of
the operator~$A$ are denoted by $A_{nm}=\langle n|A|m\rangle$ in 
this basis.
Then a matrix element of the susceptibility is given by
\begin{eqnarray}
\lefteqn{
 R(t,t_w)_{nn'}  = i \theta (t-t_w) \sum_m A_{nm}A_{mn'} } \\
&&\!\!
\times\left(e^{i(E_n-E_m)t}e^{i(E_m-E_{n'})t_w} - e^{i(E_n-E_m)t_w}e^{i(E_m-E_{n'})t}\right) \nonumber
\end{eqnarray}
The imaginary part of (\ref{Romegatw}) is 
\begin{eqnarray}
\lefteqn{
{\rm Im}\, R(\omega,t_w)_{nn'} = {\rm Re}\: \sum_m A_{nm}A_{mn'} 
\label{ImR_calc} } \\
&& \times  \int_0^{\infty}\left(e^{i\omega_{nm}\tau}-e^{i\omega_{mn'}\tau}\right)e^{i\omega_{nn'}t_w}e^{i\omega\tau} d\tau
\nonumber
\end{eqnarray}
with $\omega_{nm}\equiv E_n - E_m$.
For diagonal matrix elements $n=n'$ this implies
\begin{eqnarray}
\lefteqn{
{\rm Im}\, R(\omega,t_w)_{nn} =} 
\label{Rnn} \\
&& \frac{1}{2} 
\sum_m A_{nm} A_{mn} \left( \delta(\omega+\omega_{nm})
-\delta(\omega+\omega_{mn})\right) \ . \nonumber
\end{eqnarray}
Likewise for the correlation function
\begin{eqnarray}
\lefteqn{
C_{\{A,A\}}(\omega,t_w)_{nn'} = \sum_m A_{nm}A_{mn'} 
\label{cor_calc} } \\
&& \times  \int_0^{\infty}\left(e^{i\omega_{nm}\tau}+e^{i\omega_{mn'}\tau}\right)e^{i\omega_{nn'}t_w}\cos(\omega\tau) d\tau
\nonumber
\end{eqnarray}
and the diagonal matrix elements are
\begin{eqnarray}
\lefteqn{
C^{\rm (cum)}_{\{A,A\}}(\omega,t_w)_{nn} =} 
\label{Cnn} \\
&& \frac{1}{2} 
\sum_{m\neq n} A_{nm} A_{mn} \left( \delta(\omega+\omega_{nm})
+\delta(\omega+\omega_{mn})\right) \ . \nonumber
\end{eqnarray}
If we take $|n\rangle=|{\rm GS}\rangle$ as the ground state of our
Hamiltonian, i.e.\ the system is in equilibrium, then we know
$\omega_{nm}=E_{\rm GS}-E_m\leqslant 0$ and
$\omega_{mn}=E_m-E_{\rm GS}\geqslant 0$ for all~$m$.
For positive~$\omega$ therefore only the first terms in
(\ref{Rnn}) and (\ref{Cnn}) contribute,
and for negative~$\omega$ the second terms contribute:
this just proves the zero temperature FDT (\ref{FDT_zero}) with its sgn($\omega$)-coefficient.

Now let us assume the nonequilibrium situation described above 
where the system is prepared in some arbitrary initial state
$|{\rm NE}\rangle$ at $t=0$. One can expand $|{\rm NE}\rangle$  
in terms of the eigenstates $|n\rangle$ of the Hamiltonian
\beq
|{\rm NE}\rangle = \sum_n c_n |n\rangle 
\eeq
with suitable coefficients $c_n$.
Then the relations (\ref{ImR_calc}) and (\ref{cor_calc}) are modified like
\begin{eqnarray}
{\rm Im}\, R(\omega,t_w)_{\rm NE} &=& \sum_{m,n,n'} c^*_n c^\pdag_{n'}A_{nm}A_{mn'} \times \ldots \nonumber \\
C_{\{A,A\}}(\omega,t_w)_{\rm NE} &=& \sum_{m,n,n'} c^*_n c^\pdag_{n'}
A_{nm}A_{mn'} \times \ldots \ , \nonumber
\end{eqnarray}
where "$\ldots$" stands for the same expressions as in (\ref{ImR_calc})
and (\ref{cor_calc}). In general this will lead to a nonzero difference 
\beq
{\rm sgn(\omega)}\, C^{\rm (cum)}_{\{A,A\}}(\omega,t_w)_{NE}- {\rm Im}\, R (\omega,t_w)_{NE} \not= 0
\label{difference}
\eeq
and therefore the FDT is violated. We will next study the
violation of the FDT explicitly in the time-dependent Kondo model,
and in particular also show that the difference (\ref{difference})
vanishes exponentially fast for large waiting times~$t_w$.

\section{Time-Dependent Kondo Model}
We briefly review the results obtained in Ref.~\onlinecite{article}
for the spin dynamics of the time-dependent Kondo model.
The time-dependent Kondo model is described by the Hamiltonian
\beq
H=\sum_{k,\alpha} \epsilon^\pdag_k c^\dag_{k\alpha} c^\pdag_{k\alpha}
+\sum_i J_i \sum_{\alpha,\beta}
c^\dag_{0\alpha}\,  S^\pdag_i \, \sigma^{\alpha\beta}_i\:
c^\pdag_{0\beta} \ .
\label{Kondo_Ham}
\eeq
We allow for anisotropic couplings
$J_i=(\Jperp,\Jperp,\Jpar)$ and consider a linear dispersion
relation $\epsilon_k=v_F\,k$.
We have studied two nonequilibrium preparations in Ref.~\onlinecite{article}:
I)~The impurity spin is frozen for time $t<0$ by a large
magnetic field term $h(t)S_z$ that is switched
off at $t=0$: $h(t)\gg T_{\rm K}$ for $t<0$ and $h(t)=0$ for $t\geq 0$.
II)~The impurity spin is decoupled from the bath degrees of freedom for
time $t<0$ (like in
situation~I we assume $\langle S_z(t\leq 0)\rangle=+1/2$)
and then the coupling is switched on at $t=0$: $J_i(t)=0$ for
$t<0$ and $J_i(t)=J_i>0$ time-independent for $t\geq 0$.
For both scenarios
the impurity spin dynamics could then be described by an effective
time-dependent resonant-level Hamiltonian in terms of fermionic
solitons~$\Psi_k$ consisting of spin-density excitations:
\beq
H=\sum_k \varepsilon^\pdag_k \Psi_k^{\dag}\Psi^\pdag_k + \left\{ 
            \begin{array}{lc} \sum\limits_{kk'}
            g_{kk'}\, \Psi_k^{\dag} \Psi^\pdag_{k'} (d^{\dag}d-1/2) & , \, t<0\\
            \,&\, \\
            \sum\limits_{k} V^\pdag_k (\Psi^{\dag}_k d + d^{\dag}\Psi^\pdag_{k}) 
            & , \, t>0 
            \end{array}
            \right.
\label{effH}
\eeq
with effective parameters $g_{kk'}$ and $V_k$ (which also
depend on scenario~I or~II). 
The impurity spin $S_z$ is given by $S_z = d^{\dag}d-1/2$. Since
the effective Hamiltonian is quadratic for both negative and positive times,
it is straightforward to find an explicit solution for the impurity
orbital correlation functions and work out their dependence on
the waiting time. Detailed inspection\cite{article} shows that
the impurity spin dynamics is the same for both nonequilibrium
initial preparations I and~II.

\section{Toulouse Point}
At the Toulouse point\cite{Toulouse} with $J_{\|}/2\pi v_F=1-1/\sqrt{2}$
the mapping to the effective resonant level model (\ref{effH}) is exact
and the effective parameters $g_{kk'}$ and $V_k$ are independent of~$k,k'$.
This allowed us to express the spin-spin correlation
functions in closed analytical form\cite{article}
\begin{eqnarray} 
\lefteqn{
C^{\rm (cum)}_{\{S_z,S_z\}}(\tau,t_w)\stackrel{\rm def}{=} } \nonumber \\
&=&\frac{1}{2}\,
\langle\, \{S_z(t_w),S_z(t_w+\tau)\, \} \rangle
-\langle S_z(t_w)\rangle\,\langle S_z(t_w+\tau) \rangle
\nonumber \\
&=& \frac{1}{4}e^{-2\tau/t_{\rm B}}(1-e^{-4t_w/t_{\rm B}})\label{KondoC} \\
&&-\left(s(\tau)
-s(t_w+\tau)e^{-t_w/t_{\rm B}}+ s(t_w) e^{-(t_w+\tau)/t_{\rm B}}\right)^2
\nonumber 
\end{eqnarray}
for the symmetrized part and 
\begin{eqnarray} 
\lefteqn{
C_{[S_z,S_z]}(\tau,t_w) \stackrel{\rm def}{=} \frac{1}{2}\,
\langle\, [S_z(t_w),S_z(t_w+\tau)]\, \rangle } \nonumber  \\
&=&-i\,e^{-\tau/t_{\rm B}} \Big( s(\tau)-s(t_w+\tau)e^{-t_w/t_{\rm B}}
\nonumber \\
&&\qquad\qquad
+s(t_w)e^{-(t_w+\tau)/t_{\rm B}} \Big)
\label{KondoR}  
\end{eqnarray}
for the antisymmetrized part.
Here $s(t)\stackrel{\rm def}{=}(t_{\rm B}/\pi) \int_0^\infty d\omega\,
\sin(\omega\tau)/(1+\omega^2 t_{\rm B}^2)$ with the shorthand notation
$t_{\rm B}=\pi w t_{\rm K}$. $w=0.4128$ is the Wilson number and
$t_{\rm K}=1/T_{\rm K}$ is the Kondo time scale, i.e. the 
inverse Kondo temperature. In this paper we use the definition of the Kondo temperature~$T_{\rm K}$
via the zero temperature impurity contribution to the Sommerfeld
coefficient, $\gamma_{\rm imp}=w\pi^2/3T_{\rm K}$.

\begin{figure}
\includegraphics[width=0.45\textwidth,clip]{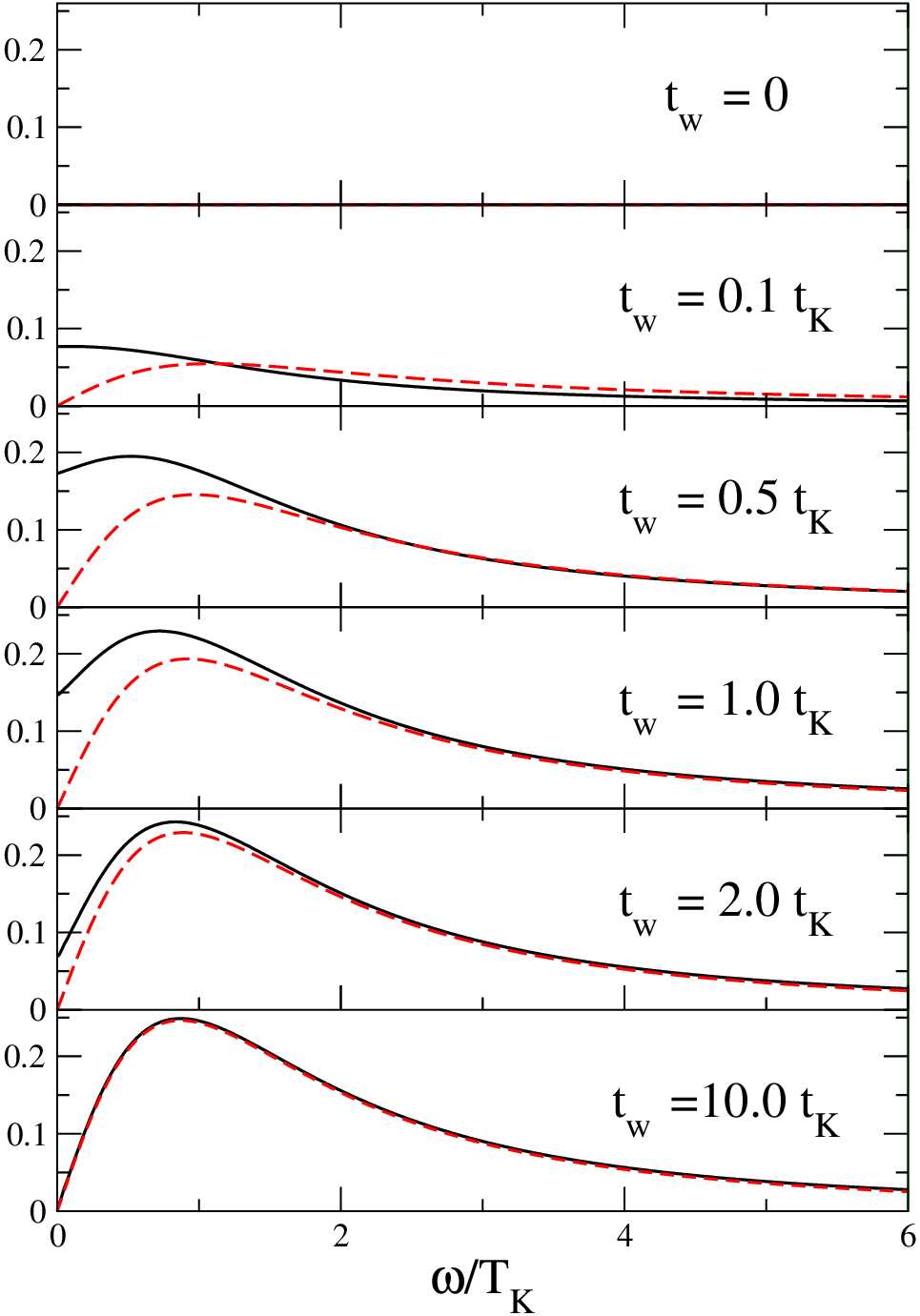}
\caption{Universal curves for the spin-spin correlation function
$T_{\rm K}\times C^{\rm (cum)}_{\{S_z,S_z\}}(\omega,t_w)$ (solid line) and response function $T_{\rm K}\times\,{\rm Im}\,R(\omega,t_w)$ (dashed line)
at the Toulouse point. Notice the normalization of the equilibrium curve
($t_w\rightarrow\infty$) which follows from (\ref{cor_sym})
with the operator identity $S_z^2=1/4$: this gives 
$\int_0^\infty C^{\rm (cum)}_{\{S_z,S_z\}}(\omega,t_w=\infty)d\omega=\pi/4$.
} 
\label{figure}
\end{figure}
{}From (\ref{KondoC}) and (\ref{KondoR}) one can obtain the Fourier transforms
(\ref{Romegatw}) and (\ref{Comegatw}). Results for 
$C^{cum}_{\{S_z,S_z\}}(\omega,t_w)$ and ${\rm Im}R(\omega,t_w)$ 
for various waiting times~$t_w$ are shown in Fig.~\ref{figure}.
For zero waiting time $t_w=0$ the FDT is trivially fulfilled since
the system is prepared in an eigenstate of~$S_z$ and therefore both
functions vanish identically,
$C^{\rm (cum)}_{\{S_z,S_z\}}(\omega,t_w=0)={\rm Im}\,R(\omega,t_w=0)=0$.
For increasing waiting time the curves start to differ, which indicates
the violation of the FDT in nonequilibrium. For large waiting time as compared to the
Kondo time scale one can then see nicely that the curves
coincide again, which shows that the system reaches equilibrium
behavior for $t_w\rightarrow\infty$ where the FDT is known to hold. 
{}From the curves in Fig.~\ref{figure} one also notices that the maximum
violation of the FDT occurs at zero frequency, while it becomes fulfilled
more rapidly at higher frequencies. We interpret this as showing that 
high-energy excitations find ``equilibrium-like" behavior faster
than low-energy excitations probed by the small~$\omega$ response.
The high-energy components of the initial nonequilibrium state 
can ``decay" more quickly for a given waiting time.

At zero frequency $C^{\rm (cum)}_{\{S_z,S_z\}}(\omega=0,t_w)$ is non-zero 
for $0<t_w<\infty$
as if one were studying the spin dynamics of the equilibrium system at
finite temperature. This leads to the definition of the {\em effective
temperature} $T_{\rm eff}$ via the zero frequency limit of (\ref{FDT})
\beq
\lim_{\omega\rightarrow 0} \frac{{\rm Im}\,R(\omega)}{\omega}
=\frac{1}{2T_{\rm eff}}\,C^{\rm (cum)}_{\{A,A\}}(\omega=0) \ .
\label{Teff}
\eeq
This concept of an ``effective temperature'' is frequently used 
and well-established in the investigation of classical
nonequilibrium systems.\cite{Kurchan}
We suggest that it is also useful in a 
quantum nonequilibrium system by 
giving a measure for the ``effective
temperature" of our bath (i.e.\ the conduction band electrons)
in the vicinity of the impurity. 

\begin{figure}[t]
\includegraphics[width=0.45\textwidth,clip]{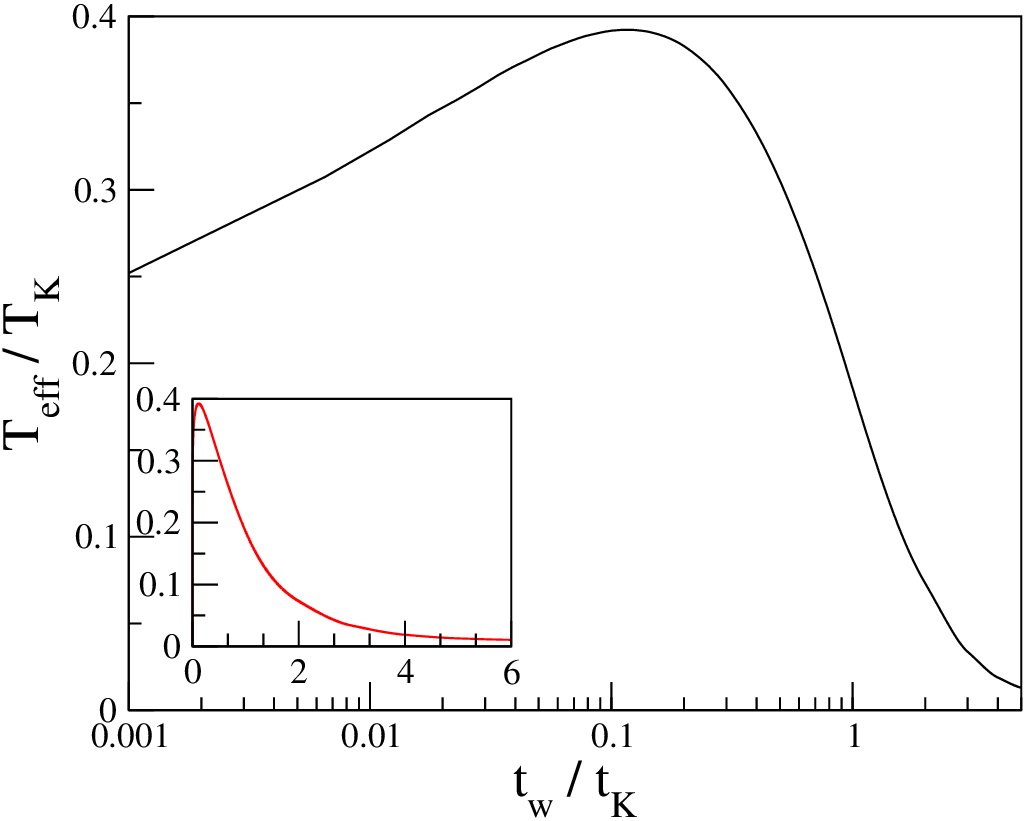}
\caption{Effective temperature $T_{\rm eff}$ 
as a function of the waiting time~$t_w$ at the Toulouse point. 
The inset shows the same curve
on a linear scale to illustrate how fast the initial heating occurs.} 
\label{figure2}
\end{figure}

We can see this explicitly by using
(\ref{Teff}) to evaluate the effective temperature as a function
of the waiting time; the             
results are shown in Fig.~\ref{figure2}. 
One sees that the effective temperature goes up very quickly as a function of
the waiting time until it reaches a maximum of $T_{\rm eff}\approx 0.4T_{\rm K}$
at $t_w\approx 0.1 t_{\rm K}$. After that the system cools down again.
We can understand this by noticing that the conduction band is initially
in its ground state with respect to the Hamiltonian for $t<0$, therefore
its effective temperature vanishes. As the spin dynamics is turned on
at $t=0$ the Kondo singlet starts building up. Its nonzero binding energy
therefore initially ``heats up" the conduction
band electrons. After a sufficiently long time the Kondo singlet has
been formed and then the process of energy diffusion takes over: the
binding energy that has initially been stored in the vicinity of the
Kondo impurity diffuses away, the system equilibrates and the effective
temperature goes back to zero. The behavior of the effective 
temperature therefore traces this competition of release of binding
energy and energy diffusion away from the impurity. Analytically
one can show for very small waiting time $t_w\ll t_{\rm K}$
\beq
\frac{T_{\rm eff}}{T_{\rm K}} \simeq \frac{1}{w}\frac{1}{\ln(\pi w\, t_{\rm K}/t_w)}
\eeq
and an exponential decay to zero temperature for long waiting time
$t_w\gg t_{\rm K}$
\beq
\frac{T_{\rm eff}}{T_{\rm K}} \propto e^{-t_w/t_{\rm K}} \ . 
\label{expdecay}
\eeq
Finally we want to emphasize that while the effective temperature seems a 
useful phenomenological concept for interpreting the $\omega=0$ behavior in nonequilibrium,
its definition (\ref{Teff}) does not capture the small~$\omega$-behavior.
Since $C^{\rm (cum)}_{\{S_z,S_z\}}(\omega,t_w)-
C^{\rm (cum)}_{\{S_z,S_z\}}(\omega=0,t_w)\propto |\omega|$ is nonanalytic 
for small $\omega$ and finite waiting time (see Fig.~\ref{figure}
and the discussion in Ref.~\onlinecite{article}),
the long time decay of the spin-spin correlation function is always
algebraic for all $t_w>0$ (therefore characteristic of equilibrium
zero temperature behavior), 
$C^{\rm (cum)}_{\{S_z,S_z\}}(\tau,t_w)\propto \tau^{-2}$.

\section{Kondo Limit}
The Kondo limit with small coupling constants $J_\perp, J_\parallel\rightarrow 0$
is the relevant regime for experiments on quantum dots. In this regime
the results in Ref.~\onlinecite{article} are not exact, but were shown to
be very accurate by comparison with asymptotically exact results
for $t_w=0$ and $\tau/t_{\rm K}\gg 1$.\cite{LesageSaleur}
For our purposes here the main difference from the
Toulouse point analysis is the nontrivial structure of the effective
parameters $g_{kk'}$ and $V_k$ from Ref.~\onlinecite{article}
in (\ref{effH}). This makes it impossible
to give closed analytic expressions like (\ref{KondoC}) or (\ref{KondoR}),
but the numerical solution of the quadratic Hamiltonian (\ref{effH})
is still straightforward. The results presented in this section were
obtained by numerical diagonalization of (\ref{effH}) with 4000~band
states. The numerical errors from the discretization are very small
(less than $2\%$ relative error in all curves).\cite{footnote_UV} 

\begin{figure}[t]
\includegraphics[width=0.45\textwidth,clip]{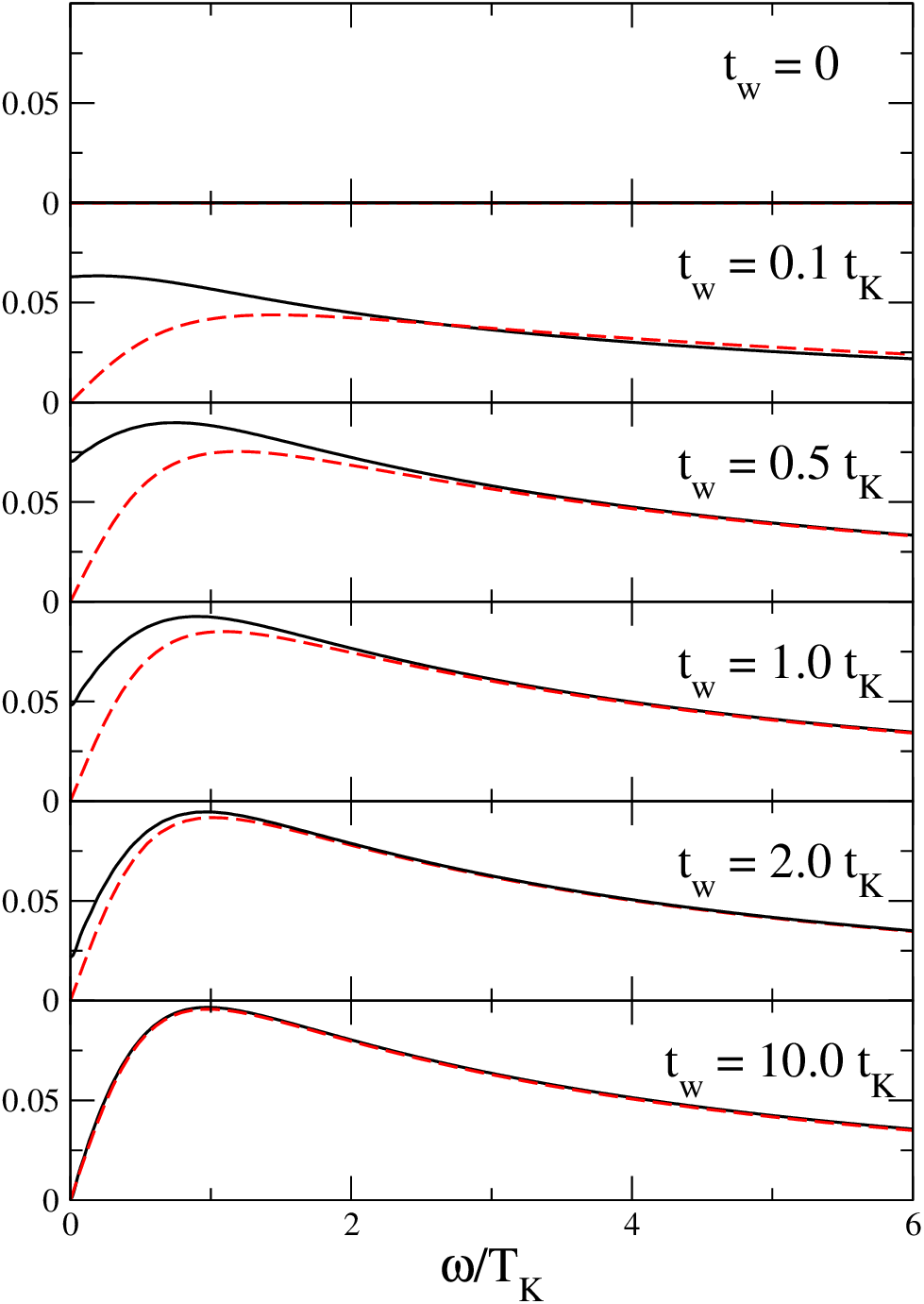}
\caption{Universal curves for the spin-spin correlation function
$T_{\rm K}\times C^{\rm (cum)}_{\{S_z,S_z\}}(\omega,t_w)$ (solid line) and response function $T_{\rm K}\times\,{\rm Im}\,R(\omega,t_w)$ (dashed line)
in the Kondo limit (compare with Fig.~\ref{figure}).}
\label{figure3}
\end{figure}

\begin{figure}[t]
\includegraphics[width=0.45\textwidth,clip]{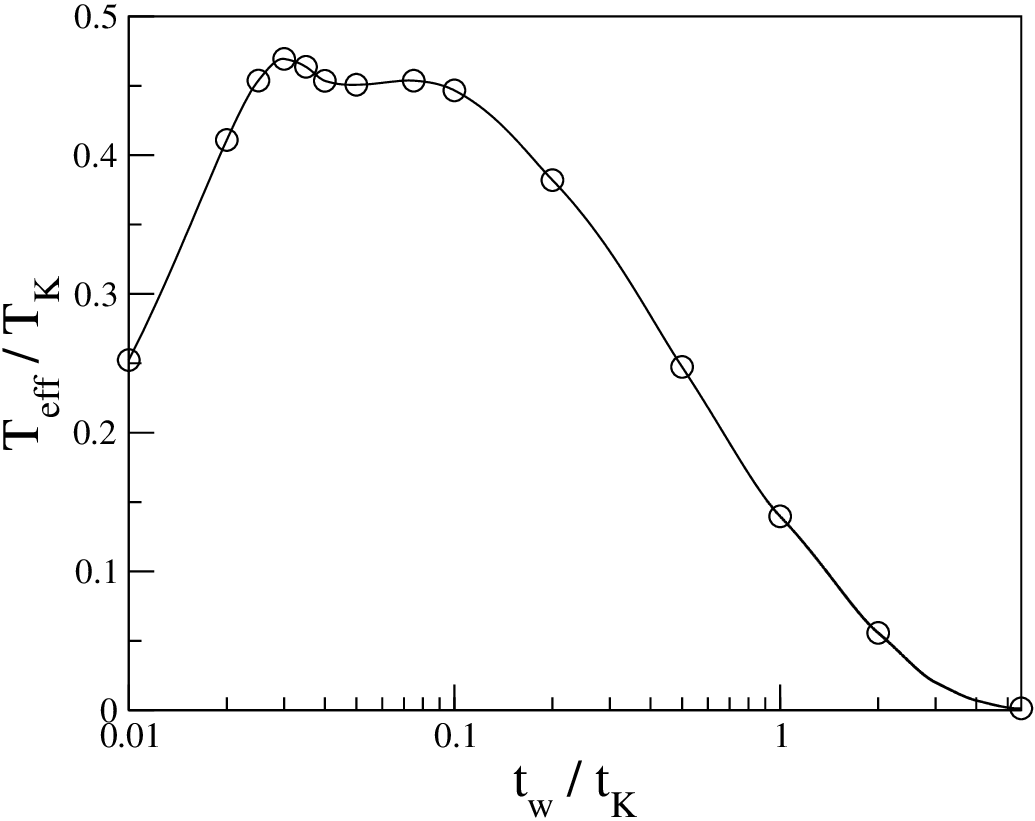}
\caption{Effective temperature $T_{\rm eff}$ 
as a function of the waiting time~$t_w$ in the Kondo limit.
The line is a guide to the eye.
The size of the datapoints (circles) indicates the numerical error.
The datapoint for $t_w/t_{\rm K}=5.5$ is numerically indistinguishable
from zero.}
\label{figure4}
\end{figure}

Fig.~\ref{figure3} shows the behavior of the spin-spin correlation
function and the response function obtained in this manner. Similar
to the Toulouse point results we observe a violation of the FDT
for finite nonzero waiting time, $0<t_w<\infty$. For $t_w\rightarrow\infty$
one recovers the FDT exponentially fast (\ref{expdecay})
as expected since the system equilibrates. 
The results for the effective temperature in the Kondo limit
are depicted in Fig.~\ref{figure4}. While the behavior of $T_{\rm eff}(t_w)$ is somehow
more complicated than at the Toulouse point, the interpretation
regarding heating and cooling effects carries over without
change. The main difference is that the maximum effective temperature
is already reached for $t_w\approx 0.03t_K$ in the Kondo limit. We
interpret this as being due to the (dimensionful) bare coupling constants at
higher energies that are larger than the renormalized low energy
scale~$T_{\rm K}$ and therefore lead to faster heating.

\section{Conclusions}
Our investigation of the zero temperature quantum limit of the 
fluctuation-dissipation theorem in the time-dependent Kondo model
provides some important lessons regarding its relevance
in quantum nonequilibrium systems. For the Kondo system prepared
in an initial state with a frozen impurity spin, i.e.\ in a {\em product
state} of system and environment, the FDT is violated for all nonzero
waiting times~$t_w$ of the first measurement after switching on the spin
dynamics at $t=0$. For large waiting times as compared to the Kondo
time scale the FDT becomes fulfilled exponentially fast, which
indicates the quantum equilibration of the Kondo system. A
quantitative measure for the violation of the FDT is provided
by the {\em effective temperature} $T_{\rm eff}$\cite{Kurchan} 
here defined via the spin dynamics (\ref{Teff}) and
depicted in Figs.~\ref{figure2} and~\ref{figure4}. It traces
the buildup of fluctuations in the conduction band:
Initially, the conduction band electrons are in equilibrium with
respect to the Hamiltonian for $t<0$. Then in the vicinity
of the impurity they get locally ``heated up"
to $T_{\rm eff}$ about $0.4 T_{\rm K}$ (Toulouse point)/
$0.45 T_{\rm K}$ (Kondo limit) 
due to the release of the binding energy when the Kondo singlet
is being formed. Eventually, this excess energy diffuses away
to infinity and $T_{\rm eff}$ reaches zero again. In this sense 
the largest deviation from zero temperature equilibrium behavior
occurs for $t_w\approx 0.1 t_{\rm K}$
at the Toulouse point, and for $t_w\approx 0.03 t_{\rm K}$ in the 
experimentally relevant Kondo limit, with very rapid initial heating  (see the inset
in Fig.~\ref{figure2}). These observations could be relevant for designing
time-dependent (functional) nanostructures with time-dependent
gate potentials\cite{Nordlander} since they give a
quantitative insight into how long one needs to wait after switching for the
system to return to (effectively) zero temperature. 

{}From a theoretical point of view it would be interesting to
study the FDT for other observables (like the current) and
in other nonequilibrium quantum impurity systems
in order to see which of the above observations are generic.
Notice that the ``effective temperature''
will generally depend on the observable chosen for its
definition in (\ref{Teff})\cite{Calabrese2}, though we suggest that the
qualitative behavior (rapid initial increase and exponential
decrease) will be similar for all local observables. 
Work along such lines is in progress in order
to substantiate the concept and notion of
an ``effective temperature'' qualitatively characterizing 
the evolving nonequilibrium
quantum state, and to explore its usefulness
in quantum nonequilibrium models in general.

\begin{acknowledgments}
The authors acknowledge valuable discussions with D.~Vollhardt. This
work was supported by SFB~484 of the Deutsche Forschungsgemeinschaft (DFG).
S.K. acknowledges support through the Heisenberg program of the DFG.
\end{acknowledgments}


\begin{thebibliography}{5}
\bibitem{footnote_UV}
Notice that a large 
number of band states is important for analyzing the behavior for
small waiting time: one must ensure that $t_w\gg 1/\Lambda$ where
$\Lambda$ is the ultraviolet cutoff in order to obtain universal
curves that only depend on the low energy scale~$T_{\rm K}$.
\end{thebibliography}
\end{document}